# The Geography of Love: Decoding the Spatial Pattern and Digital Self in Chinese Online Courtship

**Shu Wei**

*The Bartlett Centre for Advanced Spatial Analysis, University College London, UK*

ABSTRACT

The evolution of Chinese online courtship culture is a mixed product under dynamic interaction between the feudal tradition, urban development and government policies. The presented research examines the regional online dating popularity, spatial homogamy and the self-presentation pattern in the cyber love market, with a special focus on the geographical factors. More than 30 thousand recent data are collected from China's biggest online dating site jiayuan.com, and analyzed through the techniques of ordered logit modeling, Katz centrality analysis and word2Vec natural language processing (NLP). The results indicate that the high user density spots are widely spread among different parts of China rather than only clustering in the developed areas. While major users expect a within-city distance with the potential dater, the increasing social mobility and cross-city mating trend is also driven by the motivation of gaining localized access to better social and welfare resources. Moreover, people tend to emphasize the soft qualities of caring, reliable and family engagement in the online self-presentation, with the marriage pragmatism overweighing the personalized individualism and emotional self.

## 1. Introduction

The practice of everyday life today is highly entangled with digital media, and this especially extends to the social and romantic interaction (Light, 2014). Attracted by the convenient and abundant resources of access, communication, and matching services, an increasing group are turning to online dating sites in search of love and intimacy (APS, 2018). As one of the biggest online dating markets in the world, China's online dating culture yields an exclusive research value especially considering its great population, broad geographical attributes and diverse oriental culture. The current research is therefore focused on the Chinese society, and scopes down to study the online courtship, which refers to the serious relationship precedes the traditional engagement or marriage, in the cyber dating sphere (Degim, Johnson & Fu, 2015). By leveraging the spatial analysis and digital forensics expertise, it aims to discuss the following questions:

- What is the recent popularity of the online courtship among different regions in China? Does the active level of online dating activities vary from different scales of cities (Tier 1 to Tier 4)?
- What is the user requirement towards the geographical closeness of their partner? How the associated self-features and social-economic conditions affect people's mobility or their willingness to start a long-distance relationship?
- Is there any word use and semiotic patterns underlying the customized biographies in the cyber courtship sphere?

### *1.1. Ecology Matters: The Rise of Online Courtship in urbanized China*

The evolution of Chinese courtship culture in the past one century is a mixed product of the dynamic interaction between the feudal tradition, urban and economic development and government policies. Before 1950s, professional matchmaking served as the long-standing practice where a '红娘 Hong Niang' (matchmaker) introduced the single man and woman to each other under their parents' entrustment (Li & Lipscomb, 2017). Prospective matchings were selected based on the criteria '门当户对 Men Dang Hu Dui' (matching doors and parallel windows), which means the families of two sides are well-matched regarding socio-economic status (He et al., 2013). A significant progress to freedom was stimulated by the update of Chinese New

Marriage Law in 1950[1] and 1980[2] as it proposed marriage freedom by requiring full consents of each side of the marriage and permitting divorce. This destabilization of traditional courtship customs was further reinforced by the 1978 Open Door Policy[3] and the rapid domestic urbanization after (Xie, 1993). Accompanied by the affordable technology development and the prevalence of household computers, the launch of ‘jiayuan.com’ in 2003 labeled the birth of the Chinese online courtship landscape (iResearch, 2014). Seen as the taboo by the society in its starting stage with limited users, most of the flourishing online dating platforms now have great social acceptance with about 1.2 million user amounts and the estimated market size of 40.5 hundred million yuan by 2018 (iResearch, 2017).

One main drive behind the emerging popularity of Chinese online dating services is the re-boosted domestic migration. When China achieved the urbanization rate of 51.27% in 2011 with higher urban citizens than rural population (2012 China Statistical Yearbook), the domestic migrants got doubled (230 million) after 5 years' steady decline by a new pattern of 'city-to-city' migration (iResearch, 2014). While the inter-city population mobility raised a series of socioeconomic problems of employment, pension and housing, it also reformed Chinese traditional community-based networks through which people meet their spouses. Therefore, the online dating sites got increasing attention as a courtship solution for the marriageable singles to find their marital partners (Wang, Kwak & Whalen, 2014).

While an upward trend of Chinese cyber love business and its continued popularity are agreed from both academia and business perspectives (SBC, 2006; iResearch, 2017), there is an emerging debate about the future hotspots of the online dating market. The metropolis Beijing and Shanghai are believed by the most to stay as the leading player since they have the highest population of unmarried white-collar workers, internet usage and developed social infrastructure (Li & Lipscomb, 2017). However, a growing body of researchers claimed China's Tier 2 and Tier 3 cities will take the place of the dominant online dating consumers, due to its under-exploited market potential and less active offline social life (Pan, 2015). Also, the increasing returning migrants who go back to their hometown for a long-term settle down from Tier 1 cities are great likely to contribute to the online dating demographics or even become loyal users (Liu, 2018). The argument reflects the strong association of the urban ecology and online courtship popularity in the dynamic status quo, and also notes the need of further understanding the city-wise dynamics of online courtship (Wang, Kwak & Whalen, 2014).

### *1.2. Location Matters: Geographical Traits as Core Factor of Assortative Mating*

Previous works have studied the online mate selection pattern by focusing the attributes including race, gender, lifestyle and socio-economic conditions (Lardellier, 2016; Fiore & Donath, 2005; He et al., 2013), whereas limited attention is drawn to the geographical feature to understand how the users' own location and the propinquity (spatial nearness) towards others influence their online courtship behaviors. As Edwards (2013) pointed out, although the advanced transportation system benefits people with convenient daily commute and international today, people are not infinitely mobile because of the social nature of sedentary life. Therefore, the geographical closeness is believed to play a crucial role when people consider a life partner and it is likely the shorter geographical distance is preferred for the intimate relationships (Potarca, 2014). Apart from the distance variation generated by different user location, the geographical factor is also strongly associated the local language, social customs and the aesthetic appreciation of the opposite sex (Edwards, 2013). This across-factor interaction of residence location thus further impacts the way people build their standard for a well-matched partner in the cyber-romance sphere.

One recent research shed light on the effect of distance on the partner choice using the residential histories of more than 200,000 Swedish couples. Based the evidence that most couples lived within 9 kilometres of each other before moving in together, it claimed that the majority still find their partners very close by even in a modern globalized society (Haandrikman, 2018). However, Haandrikman's idea overlooked the expansion power of the internet communication which can wavier the limitation of physical space and bridge the people across regions for high-quality conversation. It is even suggested that the text-based online dating will provide an extra advantage to long-distance communicators because they can design and revise the context to deliver a socially desirable image better compared to the face-to-face acquaintance (Finkel et al., 2012). The report published by iResearch (2017) also predicted a declining importance of spatial proximity in Chinese online assortative mating. It believed the young generation will prioritize more about appearance attractiveness and personality traits, especially for those who working in the big cities having a higher socially mobility, flexibility and open-minded attitude. Nevertheless, it worth notice that most courtships are ultimately offline-oriented for the in-person interaction outside the cyber world (Finkel et al., 2012; He et al., 2013), thus further exploration is needed to understand the role of spatial homogamy through more solid examination.

[1] The New Marriage Law 1950: provided a civil registry for legal marriages, banned marriage by proxy and required both parties had to consent to a marriage

[2] Second Marriage Law 1980: officially liberalized divorce in China.

[3] Open Door policy: announced by Deng Xiaoping in 1978 to open the door to foreign businesses that wanted to set up in China.

### *1.3. Words Matters: Self-presentation in the Love Market*

To participate in the dating websites finding the potential partner, the users are required to create a personal profile which provides their demographic information, mate preference or short self-introduction on most online platforms (e.g. OkCupid.com, Jiayuan.com, eHarmony.com). This special self-presentation process attracts great attention from the sociologists and psychologists to understand the human behaviors toward achieving the strategic goal of online mating (Loa, Hsieh & Chiu, 2013; Guadagnoa, Okdie & Kruse, 2011). One common insight is that the virtual social networking increases the likelihood of deceptive self-presentation, as the online communication lacks nonverbal cues and the users can easily manipulate the presented information tailored to the desirable dates (Toma & Hancock, 2010). A study conducted in 2008 compared the presented profile photo and the real-life looks of 54 online daters, in which one-third of photos were rated as inconsistent with more females beautifying their self-portraits (Toma, Hancock & Ellison, 2008). This finding was reinforced and further developed by Xia et al. (2014) when examining the user information in Chinese dating website baihe.com. They claimed that women are more likely to lie about their physical traits such as weight and height and men are more deceptive about socio-economic capabilities like income and job titles. As Guadagnoa et al. (2013) pointed out, the patterns of how people lie in the online dating context actually reflect an image of the society and corresponding social value surrounded them, and their deceptive behaviors are more like a strategy for maximising the chance to get liked back by the crowds they like.

Seen from the above examples, most works approach the online self-presentation in dating website through either the uploaded photo (Toma, Hancock & Ellison, 2008) or numerical demographic data (Xia et al, 2014), the textual self-expression (e.g. self-introduction) comparatively received little attention. It is mainly due to the difficulty to compare and verify the online textual data towards real-life personality, since there is no clear criteria and easy-access evidence to scientifically judge about the subjective expression of each individual. In addition, the unstructured textual data have great variation in terms of format, content or even language, which makes it harder for researchers to conduct systematic measurement for both quantitative and semiotic insights. However, the freed usage of words and exclusive sentimental elements held in the biography contain rich information about the cognitive clue of the online daters' behaviors, especially when a great amount of the images and demographic data are proved to be un-honest (Toma & Hancock, 2010; Loa, Hsieh & Chiu, 2013). Therefore, it worth persistent effort to decode the online courtships through the lens of words and contextual expression, especially the optimized natural language processing (NLP) techniques are providing the growing opportunities to employ the computational analysis on the social science data (Crowston, Allen and Heckman, 2012).

## 2. Data

### *2.1. Data Sources and Processing*

To understand the modern culture of Chinese online dating and mate selection preference, the largest Chinese internet dating website jiayuan.com is chosen as the data source. The data was collected with an automated python script, which gathers 32572 user profiles registered between June to August 2018. The targeted data fields cover the information of users' basic demographic, geographic location, social-economic condition, and their requirements towards the potential dater. Additionally, the text content of their customized self-introduction is also stored for analysis, and the variable summarization can be viewed in Table 1. An extra item of proximity preference is also added by comparing the user based-location and their preferred dater location. The proximity result is formatted as a 4-scale ordinal variable, where 1 means a strict request of the same city location and 4 means the highest flexibility and broader acceptance of the potential partner's Geo-location.

| **Target Data field** | **Self-demographic** | **Social-economic condition** | **Partner requirement** | **Self-introduction** |
|---|---|---|---|---|
| **Variable** | *Numerical*:<br>Age<br>Height<br>User ID<br>*Ordinal:*<br>Education level<br>Marital status<br>*Categorical:*<br>Gender<br>*Spatial:*<br>Province<br>City | *Ordinal:*<br>Income level<br><br>*Categorical:*<br>Car ownership (yes/no)<br>House ownership (yes/no) | *Categorical:*<br>Marital status<br>"Having Profile picture?" (yes/no)<br><br>*Spatial:*<br>Ideal province or city | customized paragraph |
| **Number of variable** | n=8 | n=3 | n=3 | n=1 |

**Table. 1 - Collected data field**

For the geographical classification, the Chinese city tier system is adopted to measure and label the city based on their development level from Tier 1 (the best developed) to Tier 4 (the least developed) as figure 1.

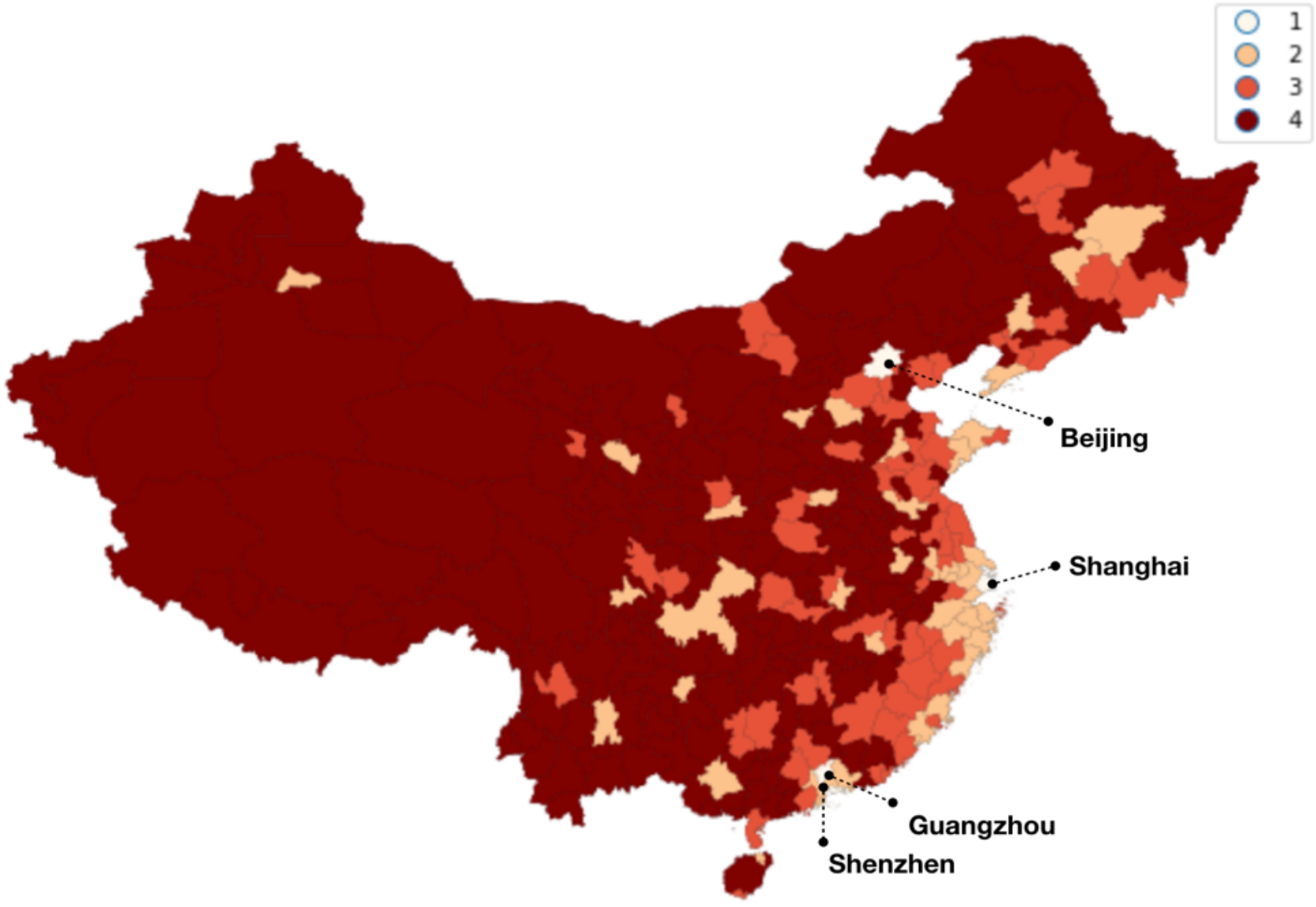

**Fig. 1 – China tier city system**

### *2.2. Collected Profiles Overview*

There is an overall number of 30754 sets of online dating profiles ($n_{male} = 19395$, $n_{female} = 11359$) retained after dropping all the data panels without user location information, since the geo-location is one main focus for the study and comparatively harder to compute for omitted data imputation.

The average age of the sample users is 30.99 for man and 29.51 for women, which are quite similar among the two gender groups. Even though the majority users (82.65%) are current single and have not married before, there is also more than 2.6% people who went through the divorce and have children with them when looking for a new partner. While half of the users (47.09%) have a bachelor degree, there is 26.51% of them who attend high school as their highest education. Apart from that, more than 1300 people (4.23%) in the sample group completed an advanced education (master or PHD). For the car and house ownership, male owners double the size of female owners in both properties (Car: $N_{male} = 510, N_{female} = 201$; House:$N_{male} = 270, N_{female} = 121$). Such pattern is also witnessed in the salary level between two genders, especially in the high-income group (monthly >20000 CNY) where the number of man (1621) is more than 2 times of woman (602). Additionally, regarding to the requirement of having a real profile photo, more male users consider it as a fixed prerequisite before they decide to go for a potential date (male: 61%; female: 49%).

In terms of geographical distribution, the users who registered and active on jiayan.com in the recent three months are based in 341 different cities across a ll the 34 provinces in China. There are 3904, 10295, 4652, 11903 users respectively in the Tier 1 to Tier 4 cities, with the gender break-down bar graph sh own as Figure 2. A density index value is calculated through normalizing the user count by the local population data ($density\ index = user\ count^2/(\frac{population}{100})$) to better compare the online dating user density. The top three cities having the highest density are Guangzhou (8.27), shanghai (8.07) and Zhengzhou (6.91), while in 5 areas (Alaer, Kizilsu, Tumushuke, BaYinGuoLeng and Shennongjia) among north west China, there is no user recorded for the chosen period.

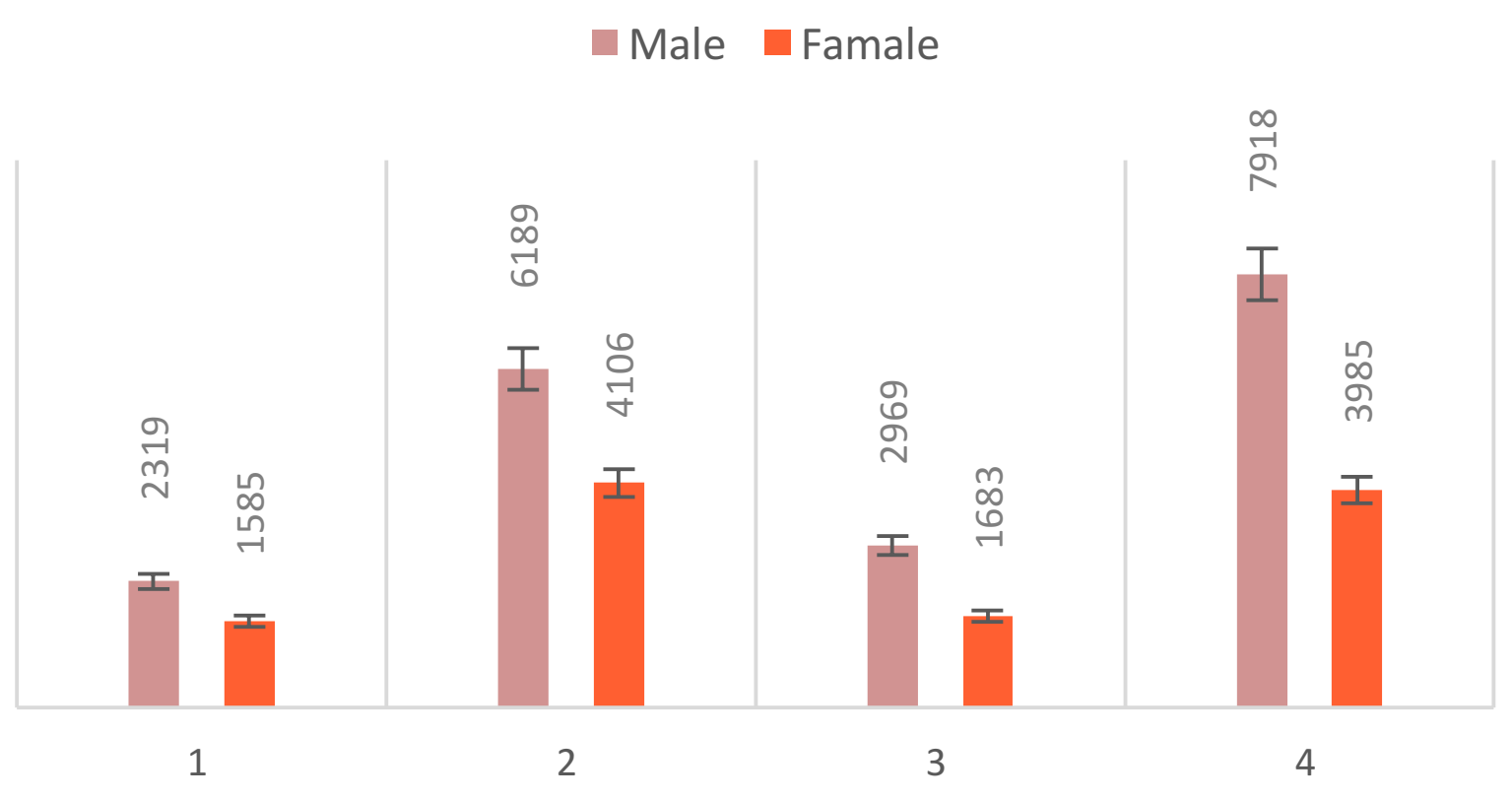

**Fig. 2 – User distribution by city level and gender**

## 3. Methodology

### *3.1. Non-parametric Testing*

The research starts to explore the data by presenting an overview of Chinese online courtship popularity among different level of cities in the past three months. A set of user density index are calculated by normalizing the regional user number against its population, and they are then also mapped to 4 ordinal categories. Considering the varied size of different city level ($N_{tier1} = 4, N_{tier2} = 43, N_{tier3} = 67, N_{tier4} = 234$), the chi-square test is chosen to detect the potential association between the city level and regional popularity of online courtship. As a non-parametric test, it examines the null hypothesis by comparing the observed and expected cell counts of online dating popularity crossable based on the formula (PSECS, 2018):

$$x^2 = \sum(O - E)^2/E \quad (1)$$

While the chi-square test only provides statistical evidence of the significance level of the potential link of two factors, a spatial visualization will be also accompanied for further sense of how the city scale affect the online dating usage in an intuitive way.

### *3.2. Ordered Logit Regression for Omitted Social Science Data*

To approach the main question of interest about the geographical-related preference of online dating users, a predictive model is built to project people's proximity preference of their potential partner. The answers towards proximity requirement here are coded to a 4-level ordinal categorical variable, where a bigger number indicating a higher flexibility or wider range of acceptance in terms of location requirement of the potential dater. The quantitative modeling is based on the ordered logit model (also known as the proportional odds model), which is considered robust to predict the probability of the users' response falling within a certain level, and widely used in social science and human behavior study (Agrawal & Schorling, 1996; Adhikari, 2015). As one of the model back up by the proportional odds assumption, it calculates the log-odds of two set of cumulative possibilities descripted in equation (2).

$$\log\left(\frac{P(Y\leq j)}{P(Y>j)}\right) = \log\left(\frac{P(Y\leq j)}{1-P(Y\leq j)}\right) = \log\left(\frac{p_1+\cdots+p_j}{p_{j+1}+\cdots+p_j}\right) \quad (2)$$

(Y=1,2,...j, and the associated probabilities are $\{p_1, p_2, ..., p_j\}$ )

Therefore, given the outcome variable having 4 levels of category, the Ordered Logit model can be written as equation (3), where α and β represent the coefficient of variable and the intercept respectively (PSECS, 2018). It is worth noted that the OL model here will return three sets of coefficients, which finalizes three equations based on the simple sequence j (Table 2).

$$p(Y_i > j) = \frac{\exp(\alpha_j + X1_i\beta1 + X2_i\beta2 + X3_i\beta3_j)}{1 + [\exp(\alpha_j + X1_i\beta1 + X2_i\beta2 + X3_i\beta3_j)]}, j = 1,2,3 \quad (3)$$

| Response of proximity requirement | Formula | Simple sequence |
| --- | --- | --- |
| **≤ 1 (1)** | $\log(\frac{p_1}{p_2+p_3+p_4})$ | 1 |
| **≤ 2 (1 or 2)** | $\log(\frac{p_1+p_2}{p_3+p_4})$ | 2 |
| **≤ 3 (1, 2, or 3)** | $\log(\frac{p_1+p_2+p_3}{p_4})$ | 3 |

**Table. 2**

The ordered logit model is run with the multiple imputation for missing data, which compute a set of plausible values to perform the statistical analysis on each individual imputed model, and then combine the multiple outputs to a better-adjusted result (Little and Rubin, 1987).

### *3.3. Network Analysis with Katz Centrality*

To assess the influence of city node in the assertive mate selection mechanism, a Katz centrality analysis is adopted for understanding spatial preference in a network effect perspective. Similar as Eigenvector measurement, rather than simply adding up the in/out flow connection for each node, the Katz centrality considers the importance level of each vertex's neighbor nodes for evaluation (Newman, 2010). This means the current network analysis gives each city vertex a score proportional to the sum of the scores of its linked cities. At the same time, Katz centrality also overcomes the limitation of Eigenvector calculation when handling the weakly-connected component of directed network by giving each node a small amount of centrality for free (Katz, 1953). Hence, each city node can have a minimum, positive centrality value transferring to other nodes by referring to them. It is generally formulated as (4), where the constant α controls the effect level and β is the bias term to avoids zero centrality values in the centrality calculation:

$$C_{Katz}(V_i) = \alpha \sum_{j=1}^{n} A_{j,i} C_{Katz}(v_j) + \beta \quad (4)$$

The people looking for a date cross-city are treated as a weighted directed network, where the user locations are represented by different nodes and their preferred mating locations serve as the outflow destinations. The edge weights are then calculated by the count of route frequency. The higher Katz centrality value is associated with a stronger influential power resulted from the higher quality inflows, which means a more favoured city by people who already based in a popular city in the mate choice list.

### *3.4. Word2Vec and Text Clustering for Natural Language Processing*

As a main prerequisite for the text clustering, the qualitative content is expected to get mapped to the quantified index. Word2Vec is one of the neural network architectures which provides the vectorised representation of the meaningful words, and it can perform the unsupervised learning to handle the unlabeled real-world data (Yi et al., 2017). The specific method type of Continuous Bag of Words (CBOW) is chosen under the Word2Vec model, which is considered suitable for dealing with large-size of data consisting of high frequency words (Huang, 2018). The general idea behind the CBOW model is to predict the probability of a specific word presence based on the given context (NSS, 2017). The implementation process is illustrated as Figure 3. All the input content with the target word are passed into the networks from the input layer, and then the average vector is calculated to become the hidden activation in the middle stage. Finally, a multiple-dimensional vector is returned, which is represented by the weight between the hidden layer and the output layer. The mathematical expression of the Continuous Bag of Words is the negative log likelihood of a word presence, therefore the possibility for the target word to appear can be written as equation (5), where $w_o$ is the output word and $w_i$ is the input word content:

$$p(w_o|w_i) = \frac{\exp(V'_{w_o}{}^{\intercal} V_{wi})}{\sum_{w=1}^{W} \exp(V'_{w}{}^{\intercal} V_{wi})} \quad (5)$$

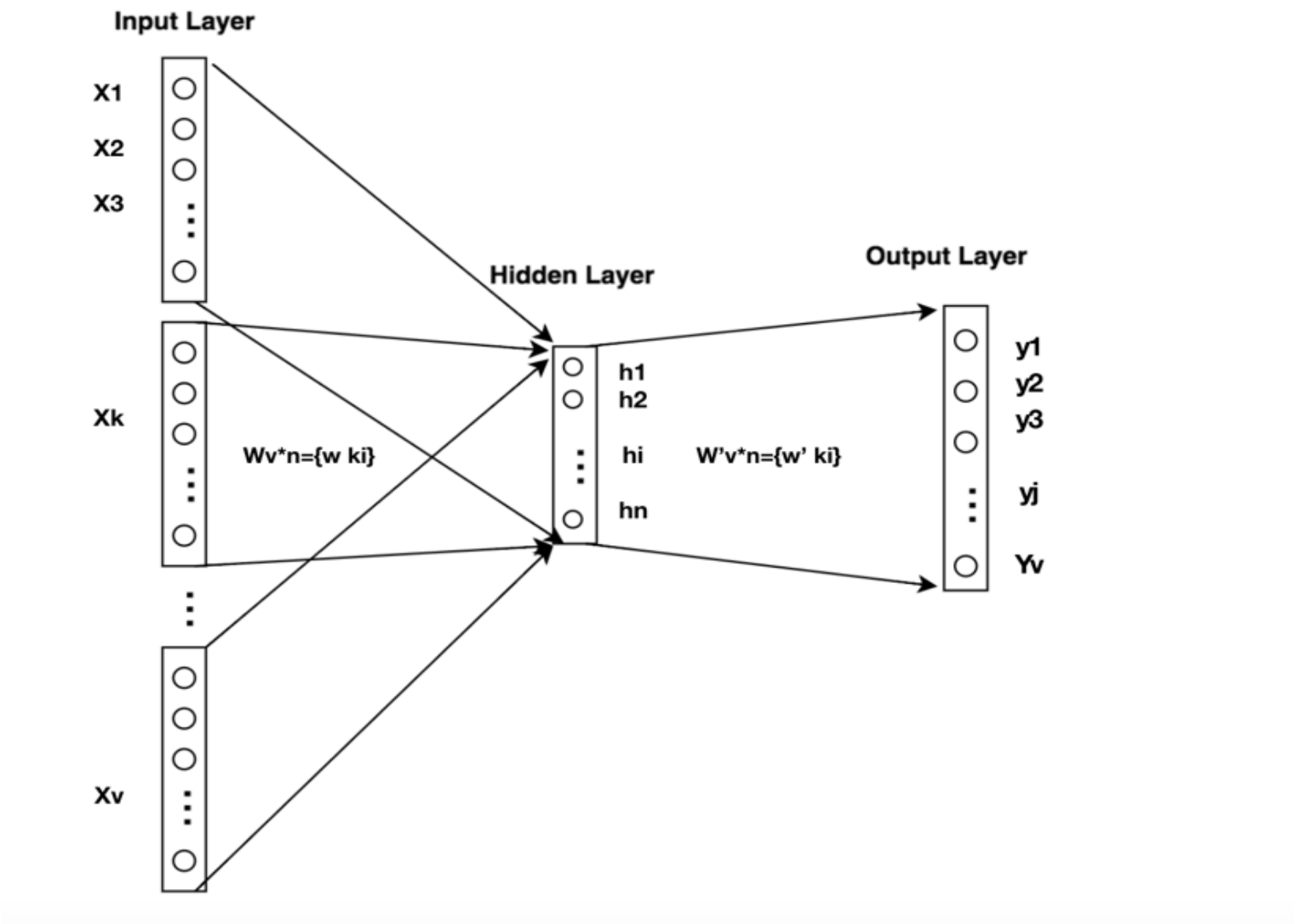

**Fig. 3 – Neural network for CBOW**

The processed word vectors are then clustered based on the hierarchical density-based method HDBSCAN[4] clustering algorithm, which takes all the vectored word list as data points and a distance metric for cluster extraction. It first estimates the density around each given data point based on the user-defined distance metric and thresholds which helps identify the noise word among the word corpus. The cluster groups are then initialized by joining together the nearby core points based on the calculated density. After that, the remaining border points of word are assigned to the suitable groups to return the final cluster results. Developed by Ester et al in 1996, the HDBSCAN clustering has considered well-suitable to handle large scale of naturally occurring data, and has been previously employed for the topic detection or sentimental grouping analysis of human generated information (Jackson, Qiao & Xing, 2018).

## 4. Data Analysis and Result

### *4.1. Online Courtship Popularity by City Level*

To explore about the relationship between city scale and online dating popularity, a cross-table is created to show the city count fallen into specific city level and online dating density level (Table 3). The Chi-square finds out that there is a strong association between the city development level (under tier city level system) and popularity of online dating usage ($x^2(4, N = 341) = 281.71, p < 0.00$).

| City count | 1 | 2 | 3 | 4 | Total |
|---|---|---|---|---|---|
| **X-high** | 3 | 1 | 0 | 0 | 4 |
| **High** | 1 | 15 | 0 | 5 | 21 |
| **Low** | 0 | 0 | 11 | 74 | 85 |
| **Medium** | 0 | 28 | 54 | 149 | 231 |
| **Total** | 4 | 44 | 65 | 228 | 341 |

**Table. 3**

Regarding to the spatial distribution of user density, compared to tier city system where the high-level cities clustering in the east coastline and southeast

[4] HDBSCAN: Hierarchical Density-Based Spatial Clustering of Applications with Noise

region of China, there is no significant clustering pattern for the online dating usage popularity (Figure 4). Even though the user density is associated with the local development condition, some tier 4 regions in south west China also have active online dating usage.

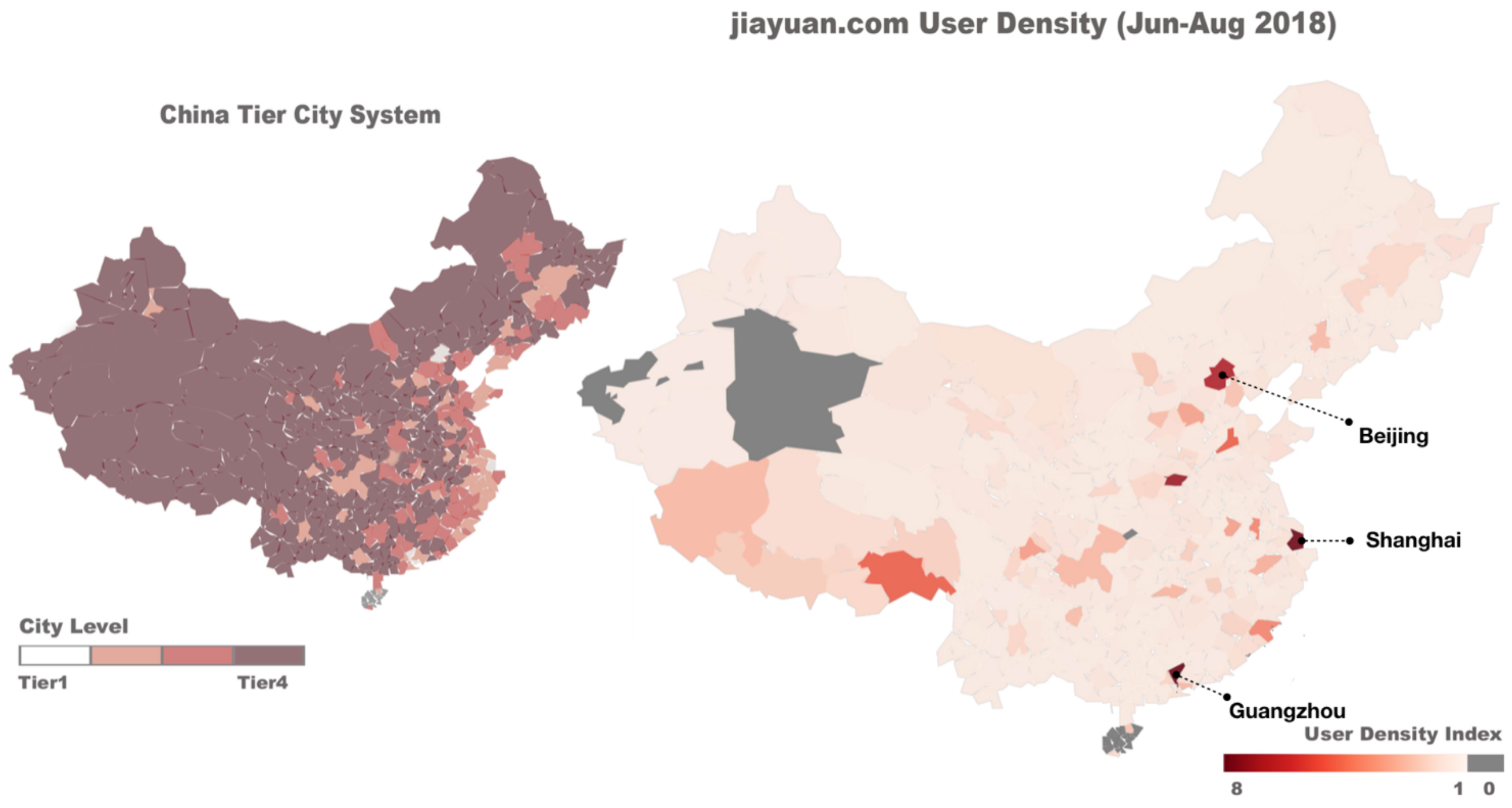

**Fig. 4 - Choropleth map of user density with the reference of Tier city system**

### *4.2. Ordinal Logit Regression for Mating Proximity Preference*

In terms of the user requirement towards the potential partner's location (proximity requirement), a distribution overview is shown as the pie chart below (Figure 5), where more than 90% people prefer the potential dater who live within the same province of them.

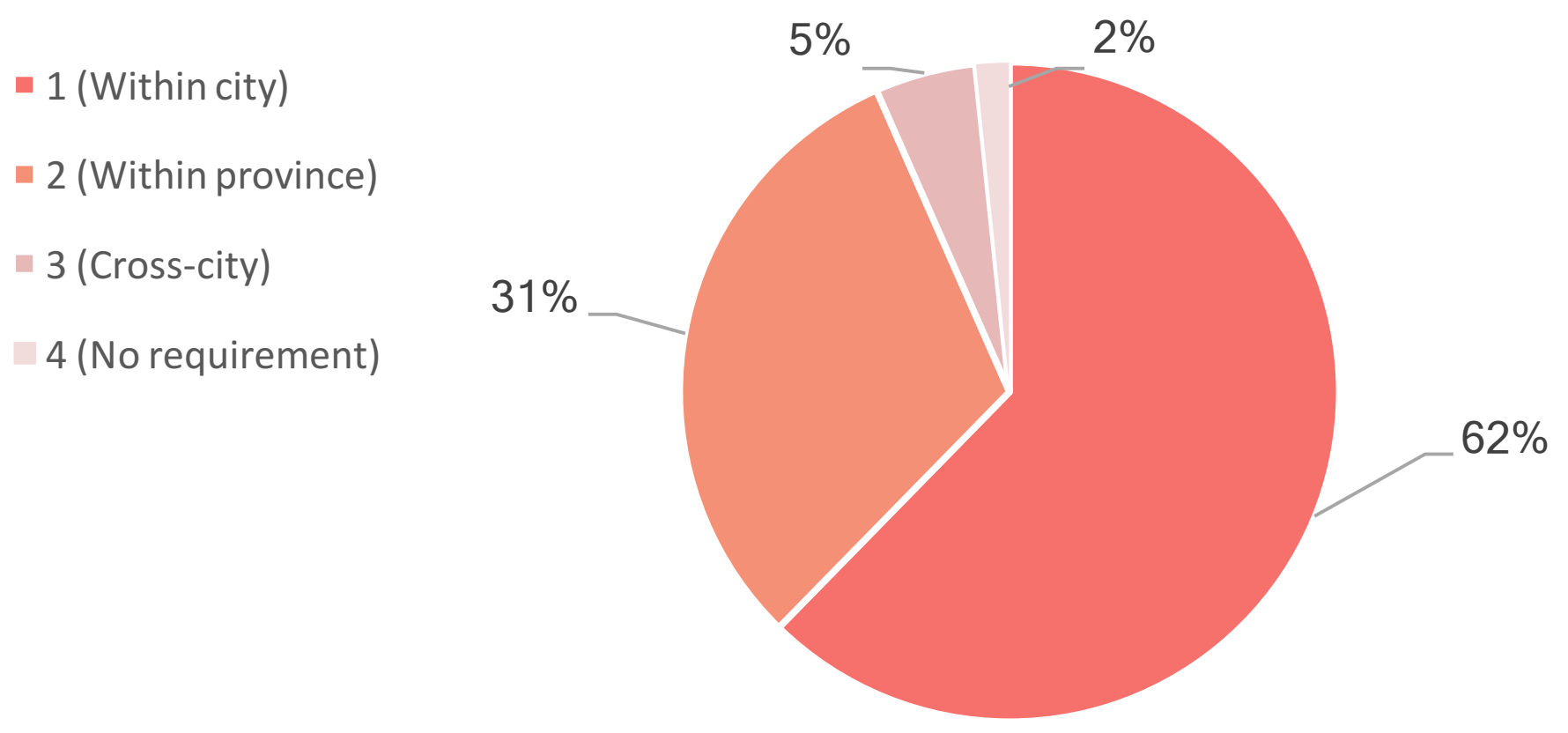

**Fig. 5 Mating proximity preference**

The ordered logit regression is run through a 10-time data imputation to model the user proximity preference by taking all the collected factors as predictor variables. An average result from the multiple data imputation is presented as table 4 and 5. It worth noticing that the items 'SalaryCoded5' and ‘EduCoded6’ are kept in the table, since all the p-value of the ordered categorical data here are computed against the baseline condition, thus a bigger p-value of them does not necessarily mean the item is insignificant (‘Gender’ is dropped as insignificant factor with p-value >0.05).

| | coefficient | Std. Error Round | t-value | p-value |
|---|---|---|---|---|
| **Age** | 0.009 | 0.002 | 5.059 | 4.22E-07 |
| **Car** | -1.15 | 0.154 | -7.449 | 9.40E-14 |
| **House** | -0.7 | 0.168 | -4.159 | 3.20E-05 |
| **cityLevelCoded2** | -0.117 | 0.047 | -2.494 | 1.26E-02 |
| **cityLevelCoded3** | 0.242 | 0.053 | 4.586 | 4.52E-06 |
| **cityLevelCoded4** | 0.348 | 0.046 | 7.582 | 3.41E-14 |
| **maritalStatusCoded2** | -0.716 | 0.043 | -16.535 | 2.05E-61 |
| **maritalStatusCoded3** | 0.394 | 0.105 | 3.745 | 1.80E-04 |
| **SalaryCoded2** | -0.268 | 0.07 | -3.858 | 1.14E-04 |
| **SalaryCoded3** | -0.645 | 0.036 | -18.058 | 6.84E-73 |
| **SalaryCoded4** | -0.392 | 0.038 | -10.295 | 7.45E-25 |
| **SalaryCoded5** | -0.001 | 0.106 | -0.005 | 9.96E-01 |
| **SalaryCoded6** | -0.75 | 0.228 | -3.294 | 9.87E-04 |
| **EduCoded2** | 0.483 | 0.035 | 13.951 | 3.11E-44 |
| **EduCoded3** | 0.874 | 0.09 | 9.7 | 3.01E-22 |
| **EduCoded4** | 0.126 | 0.039 | 3.185 | 1.45E-03 |
| **EduCoded5** | 0.77 | 0.101 | 7.619 | 2.55E-14 |
| **EduCoded6** | 0.428 | 0.233 | 1.841 | 6.57E-02 |
| **PicRCoded2** | 2.337 | 0.034 | 68.302 | 0.00E+00 |

**Table. 4 - Coefficient of all the significant explanatory variables**

| Intercepts | coefficient | Std. Error | t value |
|---|---|---|---|
| **1\|2** | 2.391 | 0.075 | 31.858 |
| **2\|3** | 4.971 | 0.08 | 62.259 |
| **3\|4** | 0.092 | 71.451 | 71.451 |

**Table. 5 - Coefficient of the response variables**

Apart from the gender, all the predictor information is significantly associated with the mating proximity requirement with the α level of 0.05. Based on the cumulative log-odd model, a positive coefficient value in Table 4 means the corresponding variable will increase the possibility of the output response falling into a lower level (a stricter requirement of the nearer distance relationship) (PSECS, 2018), and the size of coefficient indicates the impact power on the dependent variable. Therefore, in terms of the social-economic condition, when people gets more affluent, they tend to be more flexible towards the potential dater’s location, and have an increasing acceptance to far-distance relationship, while an opposite tendency is shown in the ascending education level. Additionally, regarding to the factors of age and profile picture requirement, for those who are elder or considering having the profile photo as the prerequisite of the ideal dater, they are more likely to fall into a lower level of proximity response, being less flexible about courtship mobility. However, rather than affecting the result in a consistent direction, different sub-categories under the ‘cityLevel’ and ‘martialStatus’ variable will change people’s mind-set of mating proximity requirement in a reverse way.

From the above statistical estimates of the regression output, we can also get the prediction equations as below (6), where the $y$ representing the log-odds of two cumulative possibility groups:

$$y = 0.009x_{age} - 1.15x_{car} - 0.7x_{house} - 0.117x_{cityLevel2} \dots + 0.77x_{eduLevel5} + 2.337x_{picRCoded2} \quad (6)$$

Link back to the equation (2)5 in the section 3.2, the possibility of the user's proximity preference falling into a certain level can be expressed as the following 4 equations:

$$p(1) = \frac{e^{(\alpha_{N-1}+y)}}{1+e^{(\alpha_{N-1}+y)}} = \frac{1}{1+e^{-(2.391-y)}}$$

$$p(1,\ 2) = \frac{1}{1+e^{-(4.971-y)}} \rightarrow\ p(2) = p(1,2) - p(1) = \frac{1}{1+e^{-(4.971-y)}} - \frac{1}{1+e^{-(2.391-y)}}$$

$$p(1,2,3) = \frac{1}{1+e^{-(0.092-y)}} \rightarrow\ p(3) = p(1,2,3) - p(1) - p(1,2) = \frac{1}{1+e^{-(0.092-y)}} - \frac{1}{1+e^{-(2.391-y)}} - \frac{1}{1+e^{-(4.971-y)}}$$

$$p(1,2,3,4) = 1 \ \rightarrow\ p(4) = 1 - p(1,2,3) = 1 - \frac{1}{1+e^{-(0.092-y)}}$$

### *4.3. Network Analysis of Cross-city Mating Preference*

There are 1525 users having a specific cross-city preference when they are looking for a martial partner, and these profile samples are formed as a special target group for further insights. Among all the mate selection destinations, Beijing has the highest popularity. At the same time, "Guangzhou to Beijing" (131), "Shijiazhuang to Beijing" (99) and "Taiyuan to Beijing" (43) turn out to be the top 3 popular love mobility flow in the jiayuan website (figure 6). The Katz centrality is computed with 270 city nodes in the directed network, and the 20 cities which have highest and lowest centrality value listed as Table 6.

| Top 10 | katz_centrality | City Level | Least 10 | katz_centrality | City Level |
|---:|---:|---:|---:|---:|---:|
| **Beijing** | 0.95 | 1 | Qingdao | 0.15 | 2 |
| **Shanghai** | 0.321635703 | 1 | Heilongjiang | 0.15 | - |
| **Chengdu** | 0.308706703 | 2 | Tianjin | 0.15 | 2 |
| **Chongqing** | 0.277666774 | 2 | Qinzhou | 0.15 | 4 |
| **Shenzhen** | 0.245213934 | 1 | Jian | 0.15 | 4 |
| **Changsha** | 0.243471602 | 2 | Qiandongnan | 0.15 | 4 |
| **Yantai** | 0.217448898 | 2 | Yichun | 0.15 | 4 |
| **Jinhua** | 0.216283399 | 2 | Dazhou | 0.15 | 4 |
| **Changzhou** | 0.216283399 | 2 | Guizhou | 0.15 | 4 |
| **Fuzhou** | 0.216283399 | 2 | Kelamayi | 0.15 | 4 |

**Table.6**

[5] $p(Y_i > j) = \frac{\exp\,(\alpha_j + X1_i\beta1 + X2_i\beta2 + X3_i\beta3_j)}{1+[\exp(\alpha_j + X1_i\beta1 + X2_i\beta2 + X3_i\beta3_j)]}$ (j= 1, 2, 3)

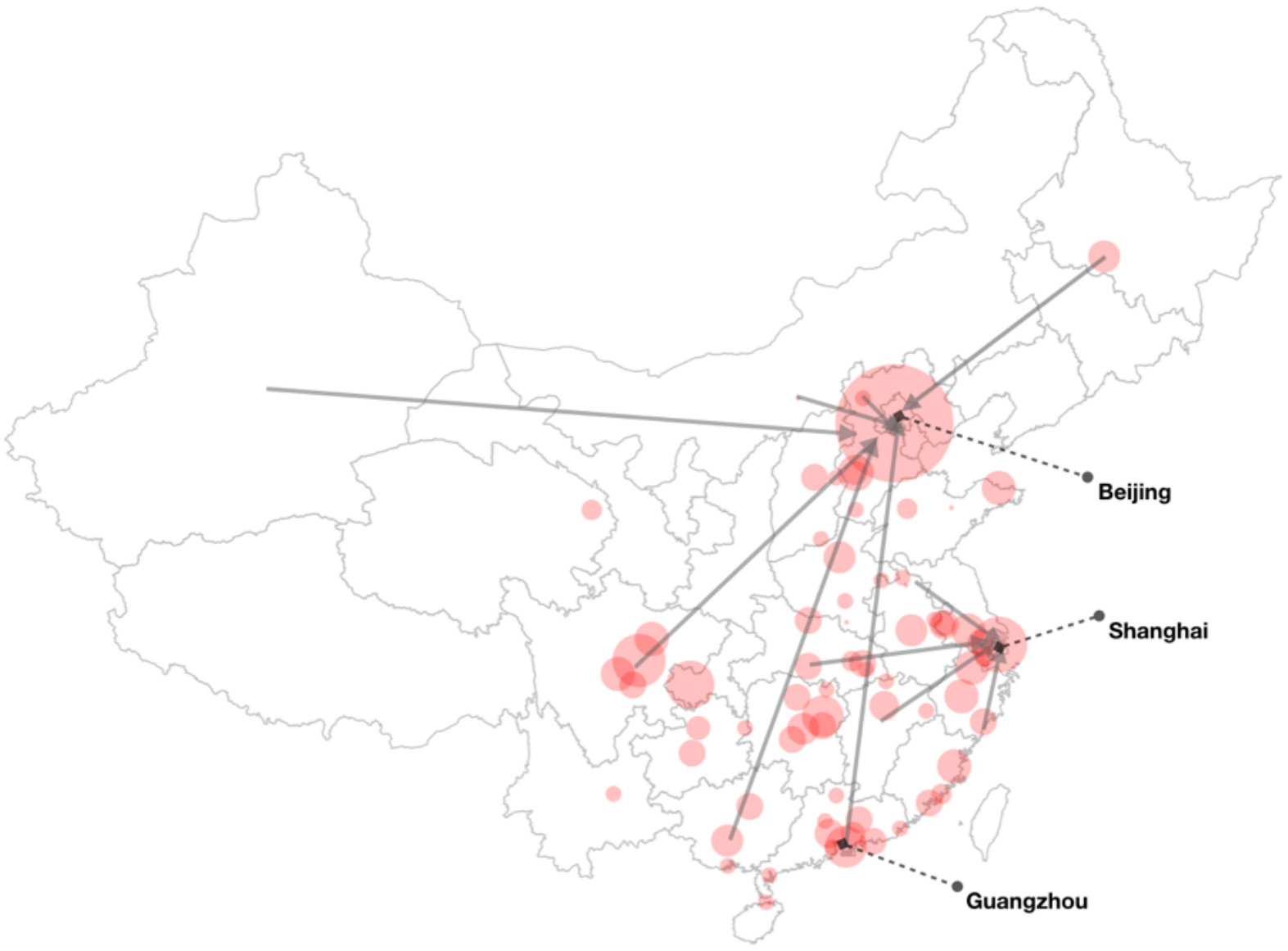

**Fig. 6 - Cross-city preference map with 10 most popular routes (circle size indicates the centrality value)**

### *4.4. Natural Language Processing with Word Clustering*

In terms of the self-introduction section in the dating profile, more than 2/3 users use the default content designed by system, and only the rest 8341 identical customized biographies are used for text analysis. These tokenized sentences return a large-size word corpus of 482172, after removing the standard stop words[6] in Chinese. A word cloud visualization is presented as Figure 7, where a larger font size indicating a higher frequency in the user contents. The most highly-used words are 爱好(hobby), 喜欢(like), 生活(life), 工作(work), 简单(simple),广泛(wide), 爱(love), 外貌(appearance), 希望(hope) and 开朗 (outgoing).

All the words in the dating profile corpus are vectorized through the word2Vec library, and the quantified word vectors make up the following 7 adjusted cluster groups (Table 7), where each group have a general topic and detected by the DBSCAN clustering method initially.

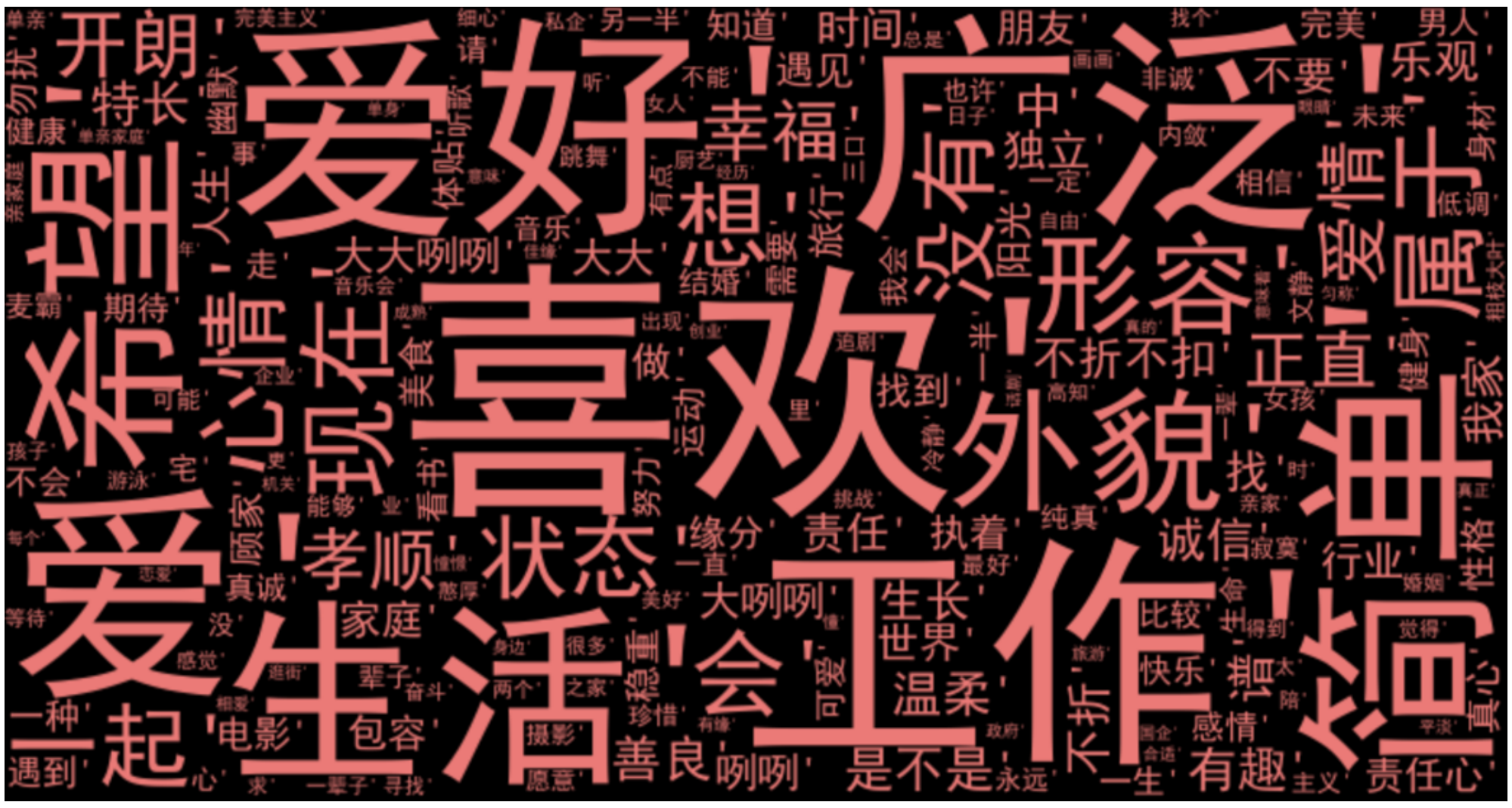

**Fig. 7 - Word cloud of the popular word corpus**

---

6 stopwords: the words which do not contain important significance to be used in Search Queries.

| **Cluster group** | **Cluster 1** |
|---|---|
| **Topic description** | Self-description of personality |
| **Key word example** | ['广泛', '简单', '心情', '孝顺', '善良', '谱', '温柔', '生长', '大大', '咧咧', '大咧咧', '大大咧咧', '遇见', '阳光', '稳重', '性格', '真心', '纯真', '细心','自由', '粗枝大叶', '高知', '亲家', '平淡', '有缘', '真的', '张扬', '小酌', '平凡', '上进心', '自我', '积极', '诚信', '责任', '比较', '宅', '麦霸', '憨厚'] |
| **Cluster group** | **Cluster 2** |
| **Topic description** | Self-description of the appearance feature (height, body shape, hair style and fashion sense etc.) |
| **Key word example** | [ '吃不胖', '胖子', '微胖', '外貌', '发福', '黑', '皮肤', '漂亮', '苗条', '高大', '白皙', '酒窝', '时尚'] |
| **Cluster group** | **Cluster 3** |
| **Topic description** | Self-description of the occupation or the industry they are working in |
| **Key word example** | ['销售', '工作', 知名企业', '完美', '完美主义', '美好', '日子', '国企', '公务员',私企', '憧憬', '自由职业', '创业', '政府', '意味着', '工作者', '金融', '外企', '晒', '脾气', '稳定', '平平', '无穷','工作狂', '高管','医务','机关'] |
| **Cluster group** | **Cluster 4** |
| **Topic description** | Hobbies in spare time |
| **Key word example** | ['生活', '希望', '旅游', '一起', '爱情', '世界', '电影', '不能', '可能', '一辈', '寻找', '出现', '厨艺', '话剧', '眼睛', '相爱', '浪漫', '瑜伽', '内心', '学霸', '人海', '伴侣', '攒钱', '追求', '每天', '音乐', '弹钢琴', '跳舞', '平平淡淡', '离开', '消遣', '游戏', '来到', '爱人', '路', '充满', '携手', '吃', '考虑', '钢琴', 'NBA', '养猫'] |
| **Cluster group** | **Cluster 5** |
| **Topic description** | Emotional terms |
| **Key word example** | ['爱好', '爱', '属于', '状态', '期待', '可爱', '努力', '我会', '相信', '合适', '身边', '谢谢', '慢慢', 'ing', '注定', '最大', '美丽', '不了', '孤独','接受' ', '了解' , '共同', '狂', '我爱你', '正在', '伤害', '共度', '太多', '改变', '享受', '用心', '能力', '渴望', '一世', '尊重', '真实', '互相', '哭', '不同', '人味', '千万', '茫茫人海', '心灵', '放弃', '不到', '宿命', '有人', '今生', '淡淡', '理想'] |
| **Cluster group** | **Cluster 6** |
| **Topic description** | Marriage view or ideal family life |
| **Key word example** | ['开朗', '幸福', '正直', '我家', '乐观', '家庭', '行业', '男人', '结婚', '也许', '事', '另一半', '一直', '主义', '寂寞', '觉得', '单亲家庭', '每个', '应该', '温暖', '认识', '开心', '网游', '对象', '付出', '之中', '方式', '三口', '温馨', '孩子'] |
| **Cluster group** | **Cluster 7** |
| **Topic description** | Slang and internet words |
| **Key word example** | ['房奴' ,'非诚', '勿扰','佳缘'] |

**Table. 7 - Word cluster of the tokenised contents**

## 5. Discussion and Conclusion

### *5.1. Love Demographics in Modern China*

Based on the larger-scale demographics from the retrieved data, the user gender ratio is 0.59 (M:11359/F:19395), which means the active male users in the website for the target period are approximately double size of the female user. This figure corresponds to the unbalanced gender odds (0.87) of 2014 Chinese Single Population Survey (iResearch, 2017), and the gap is even further amplified in the online courtship sphere, which creates an increasing invisible competition for those who are seeking an ideal girlfriend. The dominant user age range is from 26 to 34, which is consistent with the age range considered as the most suitable time to get married in Chinese society. It further proves and reinforces the value proposition of jiayuan.com to connect people for a long-term relationship and ultimately a marriage, compared to the casual dating or stranger social networking apps, which have a younger customer market segment and more diversified social functions.

The extracted profiles also sketch out the typical persona in Chinese digital mating world, where the majority was born as the only children of their family, having an undergraduate degree and working as white-collar workers in the provincial capitals (mainly tier1 and tier2 cities). The reason why online courtship is especially popular with these white-collar class is that most of them are packed with busy work that their daily exposures are mainly limited to colleagues or family. Also, they normally have a high expectation for their potential marital partner as they are the only child of their parents and carry the strong social pressure from both family and the society, where marrying well is a crucial measurement of personal success under the common social sense (Li & Lipscomb, 2017). From another perspective, it suggests that most young generation in current China have an open-minded attitude and up-to-date fashion in using social media as an effective channel for seeking the serious relationship, while online dating may be still considered taboo by the relatively conservative group (Wang, Kwak & Whalen, 2014). However, even the digital platform presents the open ideas of mate selection with great freedom, the underlying mechanism still reflects some traditional Chinese dating values from the profile questions design and aggregated user response. The material security that whether the user has a car or self-owned property is considered as the important factor of mating suitability. Additionally, for the male users, the appearance of the potential female dater is also a crucial criterion valued.

It is interesting to note that even though the regional online courtship popularity is strongly associated with the city development condition, the high user density spots are widely spread among different parts of China, rather than only clustering in the eastern developed area. An active online dating app usage is also witnessed in some regions of the tier 4 city group, which means the need of love is all around without a strict entry barrier of the social-economic development.

### *5.2. ‘An Tu Zhong Qian’ (安土重迁) VS Modern Love Mobility*

In terms of users’ preference towards the location of the potential dater, the traditional Chinese concept ‘An Tu Zhong Qian’ (安土重迁) has a strong impact on the observed response. It encourages people to be loyal and attached to one’s native land, and always reconsider the relocation decision before taking any actions. Aligned with this idea, most users have a strict proximity requirement of only considering the suitable matching within the same city as them. Also in general, people’s willingness of starting a serious relationship with someone who is physically far away are very low, since the cross-region love not only means a higher communication cost of the current interaction, but also the potential future relocation when it comes to the stage of settling down for marriage.

However, even though the nearer spatial location represents an advantage in Chinese assertive mating, there is also an emerging group showing the great flexibility and willingness to migrate or help their partner to relocate in a solid relationship. Suggested by the quantitative results of the ordered response modeling (section 4.2), the younger people tend to have a better love mobility. It could because the young generation is more comfortable with online communication for starting a long-distance relationship, and they are normally in an early stage of their career development and personal social resources accumulation, thus the potential resettlement does not weigh too strong for their romance decision. This also reflects a general trend that when people get elder, they tend to think about marriage in a more pragmatic perspective whilst the younger group are more of a romanticism (Pan, 2013). Another factor can benefit the user with a higher flexibility of proximity requirement is the increasing amount of wealth and stronger material security. In this case, the well-off people are not fussy about their dater’s location because they have more confidence in attracting the potential half to migrate to their city and help the partners to relocate using the social resources and material security they already built. Interestingly, regarding the users’ city level, people living in tier2 cities have the best love mobility or more likely to accept the partner from different regions, while those in tier1, tier3 and tier4 cities are strict about having a short physical distance with the ideal dater. The possible explanation is that people from tier2 cities have an open mindset and greater mobility due to the growing economic development and well-connected transport network, while also tend to be less picky compared to some tier1 citizens. Additionally, the different marital status also present distinguished effect on the willingness to move around for love. Compared to the single group, those who experienced a divorce are more flexible about dating someone far away from them. However, if the divorced individual also has a child with them, their love mobility suddenly drops down that they tend to only consider the daters located in the same city to avoid any potential hassles brought to their children.

Further zoom in to the special samples who have an explicit cross-city preference of partner's location, a noticeable pattern is the directed flows always point to a higher-level city compared to the origin location. It is not surprising that the premium metropoles Beijing and Shanghai are proved to be the top2 love destinations. Meanwhile, tier2 cities such as Chengdu and Chongqing are also very competitive among all the destinations, as they have a great social reputation for being the liveable cities with rapid economic development, chilling life-style and relaxed urban vibes. One strong motivation behind these cross-city preferences is to gain stable access to the better socio-economic resources in a bigger city, especially under the Chinese Hukou system. This population registration system assigns people a locational identity (either urban or rural) based on the locality of their parents, and provides people and their children generation with localised access to social and welfare resources including housing benefits, healthcare insurance and educational resources. While the urban Hukou in a big city is typically more favourable and represents higher symbolic status, it is also extremely hard to obtain, and marriage serves as a possible route to accomplish the 'identity upgrade' (Afridi, Li & Ren, 2015). Moreover, among all the resources and benefit mentioned above, the educational resource is given extreme emphasis by Chinese as it highly relates to their children growing-up environment and potential achievement. This in a way explains that Beijing shares the highest inflow and exclusive centrality attractiveness than other tier1 cities, considering it is the national capital gathering the most high-qualified teaching faculties and top-ranking institutions and universities.

### *5.3. Deep-dive in the Digital-self Behind the Words*

While the numerical figure and categorical answers sketch out the general demographics of the online love seekers, the customized word use and associated sentiment reveal more about tangible façade embedded in the digital-self. As is suggested in the frequency-based word cloud (fig.7), people generally associate themselves with a dominant positive personality by adding the popular features such as passionate, sincere, thoughtful and responsible. Compared to the other social network such as Instagram and Weibo, where the young generation make great attempt to build up a unique persona and label the self being different or even odd (Guadagno, Okdie & Kruse, 2012), the users exert more emphasis on the compatibility of interpersonal interaction and family engagement in this marriage-focused virtual community. Moreover, rather than the strong self-disclosure of personal emotion or sentimental stories, people in jiayuan.com tend to gentle their sensitivity and emotions to present a mild version of the self.

Through the lens of appearance self-description, we also sense about the common aesthetic standard towards different gender under the theme of marriage. It is noted that female users are more focused on their facial features description by specifying the detailed treats like '大眼睛' (big eyes), '白皙' (light and delicate skin) and '小酒窝 ' (small dimple), and aim to build a sweet and considerate image from how they look. Comparatively, the men give higher priority to their body shape, and using '高大' (tall and strong) and '精壮' (muscular) as the most popular labels to promote their manhood. These visual impression patterns are further reinforced by their presented hobbies. While the spare-time activities like reading, yoga, baking and playing piano are highly mentioned by women, the top items among male users are work-out, gaming and traveling. A significant gender-based persona construction is seen in the above self-presentation, and it also aligns with the mainstream stereotypes in Chinese society that the masculinity is associated with qualifies such as 'adventure', 'toughness in mind and body' and 'autonomy', while females are expected to be caring, nurturing and making time for families (Chen, 2018).

However, the quality differentiation from gender gets blurred regarding to the communication of marriage view, especially for guys. One example is many men here claim that they would like to take daily care of their future child as well as being in charge of the household chores, and mention their nice cooking skills as a side-note evidence. These single men associate themselves more with soft feminine qualities to show they are caring and trustworthy as ideal husband material. Meanwhile, female users continuously stress on their patience and supportiveness to bond the family in harmony. It is worth noticing that both sides approach the topic of relationship/marriage view from the perspectives of the family commitment, child raising and parents caring, which are concrete and pragmatic. This is different from the thinking-styles seen in the most western dating websites (e.g. EHarmony, OkCupid), where people mention more about chemistry, romance, mutual understanding between only two of the subjects (Guadagno, Okdie & Kruse, 2012). Therefore, a marriage-oriented and family-prioritized mindset serves as a core skeleton in the online love presentation, while the emotion-focused individualism is partly overlooked in the mediated courtship sphere.

### *5.4. Conclusion*

The presented human geography research has sampled the online courtship culture, one of the most vivid parts of the social structure in modern China, to explore the complexed interaction between mate selection, digital technology and urban development. The findings are interesting and thought-provoking. Although the regional online courtship popularity is strongly associated with the city development condition, the high user density spots are actually widely spread among different parts of China rather than only clustering in the developed districts. While 'the nearer the better' serves as a general attitude for the proximity preference, there is still a growing group showing better social mobility and wider geographical acceptance, especially the people from tier 2 city, and this preference can be greatly affected by the personal demographic and social-economic treats. For the cross-city mate selection, the outflows are generally aiming to a bigger and more developed destination. The main drive is to gain localized access to the social and welfare resources, including housing benefits, healthcare insurance and educational resources for their future children. For the textual self-presentation, people consciously build up the image which panders to the common aesthetic standard and gender stereotypes of the society, while both men and women highly emphasize the soft qualities of caring, reliable and family engagement. The marriage pragmatism overweighs the personalized individualism and emotional self. Future work can be added by further diving into the user interaction network and messaging behaviors in the virtual courtship community to gain deeper insights into their online behaviors. Additionally, it will be interesting to conduct the follow-up study to see the long-term life satisfaction of the matched couples from different regions, adding the time-series element to the empirical human geography research. （5982 words）